\documentclass[5p,twocolumn]{elsarticle}
\usepackage{lineno,hyperref}
\usepackage{color}
\usepackage{graphicx}

\makeatletter
\def\ps@pprintTitle{%
   \let\@oddhead\@empty
   \let\@evenhead\@empty
   \let\@oddfoot\@empty
   \let\@evenfoot\@oddfoot
}
\makeatother
\biboptions{sort&compress}
\begin{document}

\begin{frontmatter}

\title{Performance study and calibration strategy of the HADES scintillator TOF Wall with fast digital readout %\tnoteref{mytitlenote}
}

\author[tud,wut]{G. Kornakov\corref{cor1}}\ead{georgy.kornakov@cern.ch}
\author[rez,a]{L. Chlad}
\author[tum]{J.~Friese}
\author[tud,gsi]{T. Galatyuk}
\author[rez]{A. Kugler}
\author[gsi]{J. Markert}
\author[fra]{J. Michel}
\author[gsi]{J. Pietraszko}
\author[rez]{O. Svoboda}
\author[rez]{P. Tlusty}
\author[gsi]{M. Traxler}
%\author[gsi]{An. Author\corref{cor1}\fnref{fn1}}


%\ead{author2@mail.com}

\cortext[cor1]{Corresponding author}
%\cortext[cor2]{Principal corresponding author}
\address[tud]{Technische Universit\"at Darmstadt,  Institut f\"ur Kernphysik, Schlossgartenstr. 9, 64289 Darmstadt, Germany}
\address[rez]{Nuclear Physics Institute of the CAS, \v{R}e\v{z} 130, 250 68, Czech Republic}
\address[tum]{Physik Department E62, Technische Universit\"{a}t M\"{u}nchen, 85748~Garching, Germany}
\address[gsi]{GSI Helmholtzzentrum f\"ur Schwerionenforschung GmbH. Planckstra{\ss}e 1 64291 Darmstadt Germany}
\address[fra]{Goethe Universit\"at, Institut f\"ur Kernphysik, Max-von-Laue-Str. 1, 
60438 Frankfurt am Main, Germany}
\address[wut]{Warsaw University of Technology, Wydzia\l{} Fizyki, ul. Koszykowa 75, 00-662 Warszawa, Poland }
\address[a]{also at Charles University, Faculty of Mathematics and Physics, 12116
Prague, Czech Republic}

\begin{abstract}
We present in this work the calibration procedure and a performance study of long scintillator bars used for the time-of-flight (TOF) measurement in the HADES experiment. 
The digital front-end electronics installed at the TOF detector required to develop novel calibration methods. 
The exceptional performance of the spectrometer for particle identification and pointing accuracy allows one to determine in great detail the response of scintillators to minimum ionizing particles. 
A substantial position sensitivity of the calibration parameters has been found, in particular for the signal time walk. 
After including the position dependence, the timing accuracy for minimum ionizing particles was improved from 190~ps to 135~ps for the shortest rods (1475 mm) and to 165~ps for the longest (2356 mm). 
These results are in accordance with the time degradation length of the scintillator bars, as determined from previous measurements.   
\end{abstract}

\begin{keyword}Fast plastic scintillators\sep Time-of-flight\sep Heavy-ion collisions \sep time-walk correction
\end{keyword}

\end{frontmatter}

%\linenumbers

\section{Introduction}

One established technique for particle identification (PID) in nuclear and particle physics is the time-of-flight method. 
It allows one to discriminate between different particle species of reaction products using the momentum--velocity and energy-loss--velocity correlations. 
The better these quantities are known, the broader is the energy range where the method is applicable, usually up to particle momenta of several GeV$/c$.
The velocity is reconstructed as the ratio between the particle's flight path length, $D$, and the elapsed time, $T$. 
Detectors used for this purpose have a fast signal response. In a number of  experiments~\cite{Tsujita:1996tk,Kichimi:2000uu,Denisov:GlueX_position_dependent_walk_2002,Agodi:Hades_TOF_2002,Wu:2005xk,Moskal:2014rja,Carman:2020fsv} plastic scintillators are chosen for this task.   

The upgrade of the High Acceptance DiElectron Spectrometer HADES \cite{Agakishiev:Hades_Overwiew_2009}, operated at GSI Helmholtzzentrum in  Darmstadt, demanded the replacement of the readout chain for the phototube readout of the scintillator time-of-flight wall. 
The front-end electronics (FEE) part of this chain consisted of a constant fraction discriminator (CFD) plus a time to digital converter (TDC) and analogue to digital converter (ADC) branches as described in Ref.~\cite{Agodi:Hades_TOF_2002}. 
For the new readout, a significantly faster FEE, based on TRB boards~\cite{TRBCollaboration}, has been developed and integrated into the data acquisition scheme. 
The new scheme requires that the time and pulse amplitude information have to be stored into one single digital pulse. 
In such a concept, the time is encoded in the leading edge of the digital pulse and the amplitude in its width using the time over threshold (ToT) method. 

As a consequence of having the fixed-threshold discriminator, a correlation appears between the pulse amplitude and the time the signal crosses the fixed voltage threshold (walk effect). Moreover, this correlation turned out to be dependent on the particle impact position along the scintillator rod. 
Previously, with the CFD chain, the time walk correction was not necessary~\cite{Agodi:Hades_TOF_2002}.  
The time shift typically follows a characteristic exponential or power law behaviour with respect to the pulse amplitude~\cite{Tsujita:1996tk,Kichimi:2000uu,Wu:2005xk,Kim:2015tva,Denniston:2020gmc,Carman:2020fsv}.
In most of the cases, it was found that the measured signal amplitude is proportional to the number of photons arriving at the photon detectors. In these cases, a single correction factor is sufficient for a good timing reconstruction.
However, the additional impact position dependence of the time walk effect requires more refined corrections. 

Position-dependent correction constants have already been introduced for the counters of the Belle TOF~\cite{Kichimi:2000uu}, BESIII TOF ~\cite{Wu:2005xk}, GlueX setup~\cite{Denisov:GlueX_position_dependent_walk_2002} and CLAS-12 Forward TOF~\cite{Carman:2020yma}. In these cases, the residual correlation between the time and the position of the rod has been corrected using a polynomial function.
Alternatively, the position-dependent time-walk correction can be resolved using the signal template method~\cite{Moskal:2014rja,Moskal:2020jrl}. This approach resolves the systematic offset using a library of prerecorded signals that are compared to the measurement. 
Afterwards, the best fitting template is used to obtain the correct hit time and charge. 
However, this last technique cannot be applied here as the signal is stored in a single digital pulse and the polynomial offset method was not sufficient.   

Hence, an alternative calibration procedure for the TOF wall has been developed. 
The timing performance of the detector has been studied in detail with respect to the particle's deposited energy loss and the light trajectory length inside the scintillator rods. 
As a result of the calibration, the timing performance improved significantly.    

This paper is organized as follows: in Sec.~\ref{sec_exp} the experimental set-up and the observables are introduced, followed by a description of the calibration procedure in Sec.~\ref{sec_method}. 
The timing performance of the detector is then presented in Sec.~\ref{sec:performance}. Finally, the results and methods are summarized and discussed. 

\section{Experiment}
\label{sec_exp}

The data used for this analysis have been recorded with HADES in a fixed target configuration for central Au+Au collisions at $\sqrt{s_{NN}}=2.42$~GeV with a reaction trigger setting which corresponds to semi-central collisions ($\sigma = 0 - 0.43 \cdot \sigma_{\rm tot}$), covering the innermost impact parameters \cite{Adamczewski-Musch:2017sdk}.

\begin{figure}
	\centering
	\includegraphics[width=0.45\textwidth]{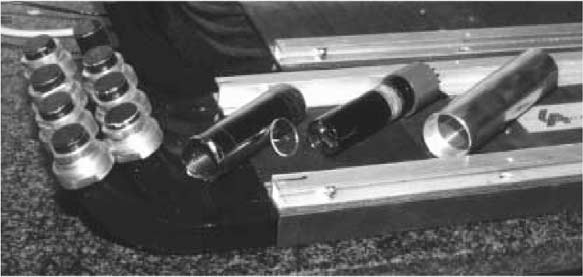}
	\caption{\label{fig_picture_one_module_8_rods} Close view of a 8-rod scintillator module and lightguides, a magnetic shield, a silicon disk for optical coupling, a  photomultiplier and the light-tight canning system.}
\end{figure}

The TOF detector in the HADES experiment consists of two different subsystems. 
A fast segmented multigap resistive plate chamber detector (RPC)~\cite{Kornakov:2014cua} covers the forward polar angles from 15$^\circ$ to 45$^\circ$. 
A wall of plastic scintillator rods~\cite{Agodi:Hades_TOF_2002} covers the polar angle range from $\simeq 45^\circ$ to $\simeq 85^\circ$ with azimuthal segmentation into 6 equal sectors. 
Each sector is equipped with 8 scintillator modules of different lengths and cross sections. 
One module comprises 8 equal rods coupled to photomultipliers (PMT) (EMI 9133B with 1'' diameter) on each side by light guides as shown in Figure~\ref{fig_picture_one_module_8_rods}. 
The scintillator material is BC408 and was manufactured by BICRON\textsuperscript{TM}\footnote{BICRON, 12345 Kinsman Road, Newbury, OH 44065, USA}. 
The dimensions of the rods are listed in Table~\ref{table_tof_geometry}. 
The read-out of the signal is done by splitting the pulses into slow and fast components for amplitude and time measurements, respectively, that are individually discriminated within a NINO chip \cite{Anghinolfia:NINO2004} and combined into a single digital pulse that is sent into a general purpose timing and readout board~\cite{TRBCollaboration} which provides a precise time digitization. 
The leading edge and width of the pulse encode both the time ($T$) and ToT, respectively. 
The ToT is converted into pulse amplitude (Q) by a two-slope linear function which linearizes the measured quantity. 

\begin{table}
\centering
	\caption{\label{table_tof_geometry} Geometry of TOF detector modules. Each module comprises 8 equally long scintillator bars instrumented at both sides with PMTs.}
	\begin{tabular}{ ccc }
		\hline 
		Module & Length (mm) & Cross section (mm$^2$) \\ 
		\hline 
		1& 2365 & 30$\times$30 \\ 
		2& 2265 & 30$\times$30 \\ 
		3& 2135 & 30$\times$30 \\ 
		4& 1970 & 30$\times$30 \\ 
		5& 1940 & 20$\times$20 \\ 
		6& 1795 & 20$\times$20 \\ 
		7& 1625 & 20$\times$20 \\ 
		8& 1475 & 20$\times$20 \\ 
		\hline 
	\end{tabular} 
\end{table}

\begin{figure}
	\centering
	\includegraphics[width=0.45\textwidth]{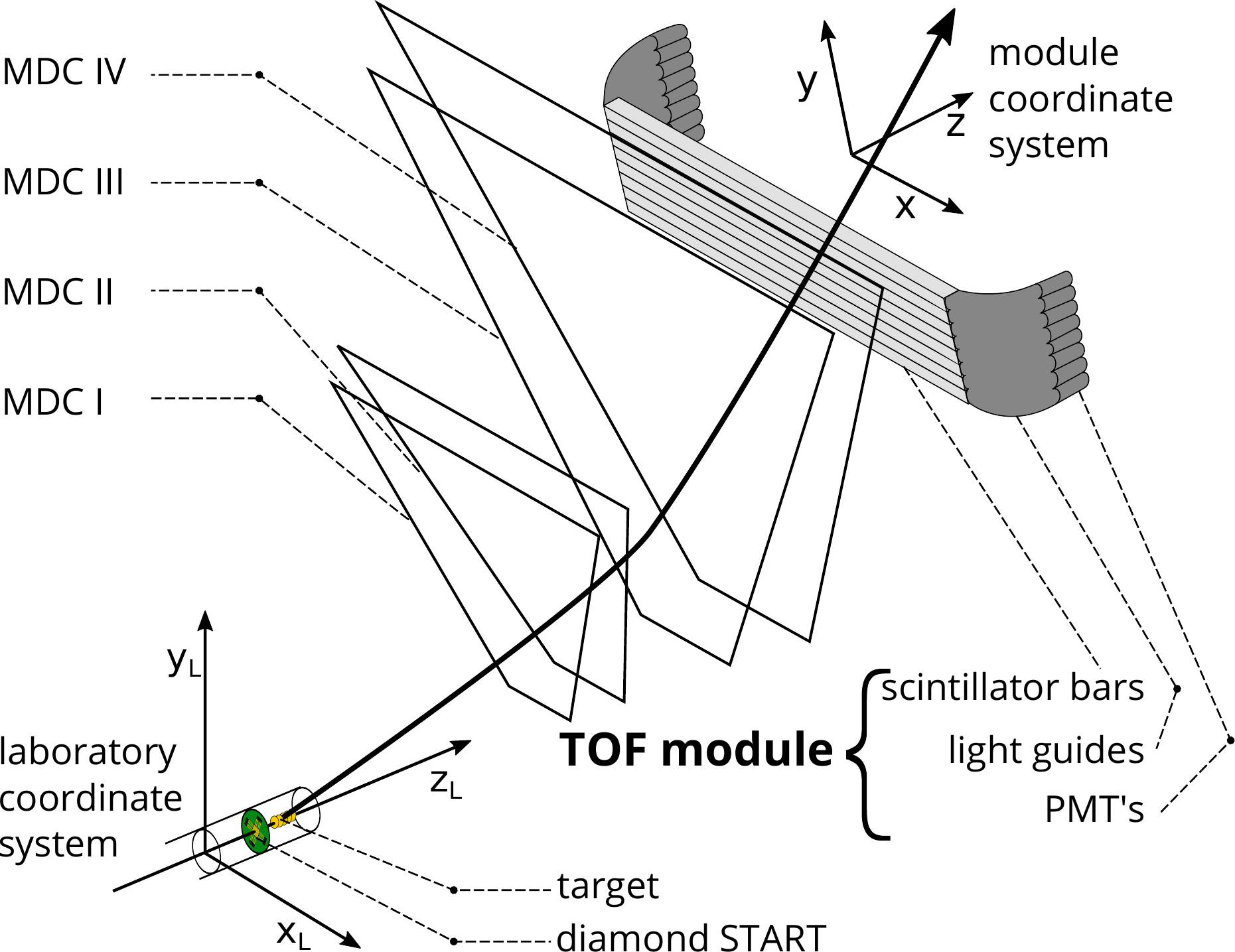}
	\caption{\label{fig_one_sector_layout} Schematic view of a particle trajectory in HADES from the target to a TOF wall scintillator. The relevant coordinate systems are shown for completeness.}
\end{figure}

The trajectory of a charged particle is schematically depicted in Figure ~\ref{fig_one_sector_layout} together with the relevant coordinate systems. 
A fast diamond in beam $T_0$ detector (START) close to the target provides an interaction time signal of the collision \cite{START}. 
The momentum and the flight distance from the target to the TOF Wall are measured by four mini drift chambers (MDC) tracking stations, two in front and two behind the superconducting toroidal magnet (for further details see ref.~\cite{Agakishiev:Hades_Overwiew_2009}). 
Each track is propagated to the middle plane of the scintillator. 
The accuracy of the projection in the TOF Wall module x-coordinate is about 1~mm. 
The flight path is reconstructed using a Runge-Kutta filter. 
The average distance between the target and the TOF Wall is about 2100~mm.
The velocity of the particle is calculated if a spatially coincident hit is reconstructed in the TOF Wall.

The hit position in the TOF detector is determined from the rod number (y-coordinate) and the x-coordinate of a rod. 
The latter is obtained from the differences of the corrected times $T_{\rm L}$ and $T_{\rm R}$ measured at the left and right sides, respectively, multiplied by the light group velocity in the scintillator, $v_{\rm g}$:
\begin{equation}
x = \frac{1}{2}(T_{\rm R}-T_{\rm L})\times v_{\rm g}.\label{eq:x}
\end{equation}
The time of flight is calculated averaging the left and right times minus the interaction time, or START ($T_{\rm S}$), 
\begin{equation}
T = \frac{1}{2}(T_{\rm L}+T_{\rm R}) - T_{\rm S}.\label{eq:tof}
\end{equation}
An alternative method is to obtain the weighted average of left and right times $T_{\rm WA}$~\cite{Denisov:GlueX_position_dependent_walk_2002} as 
\begin{equation}
    T_{\rm WA} = \frac{{T_{\rm R}}{w_{\rm R}^2} +{T_{\rm L}}{w_{\rm L}^2}  }{w_{\rm R}^2+w_{\rm L}^2} - T_{\rm S},\label{eq:tof_weighted}
\end{equation}
where the weights $w^2_{\rm L/R}$ are obtained from the inverse of the variance $\sigma^2_{\rm L/R}(x)$. Following previous work~\cite{Kurata:time_degradation_1994}, the uncertainty of the time measurement as a function of the position $x$ in the scintillator rod of length $L$ is defined as 
\begin{equation}
    \sigma_{\rm L/R} = C_{\rm L/R} e^{(x-L/2)/\lambda_{\rm D}},\label{eq:resolution_left_right_time} 
\end{equation}
where $\lambda_{\rm D}$ is the precision degradation length of the timing signal and $C_{\rm L/R}$ is a constant. 
It is important to notice that the alternative method relies on an external measurement of the intersection of the particle trajectory and the scintillator and cannot be applied to stand-alone detectors.  

The deposited energy $E_{\rm dep}$ is extracted from the geometric mean of the signal amplitudes on both sides $Q_{\rm R}$ and $Q_{\rm L}$ as 
\begin{eqnarray}
E_{\rm dep} = ke^{L/2\lambda_{\rm att}}\times\sqrt{Q_{\rm L} Q_{\rm R}},\label{eq:elos}
\end{eqnarray}
where $k$ is a free normalization parameter which translates the measured energy into units of minimum ionizing particles (MIP) and  $\lambda_{\rm att}$ is the effective optical attenuation length.

\section{Calibration methods}
\label{sec_method}

\begin{figure}[tb!]
	\centering
    \includegraphics[width=0.47\textwidth]{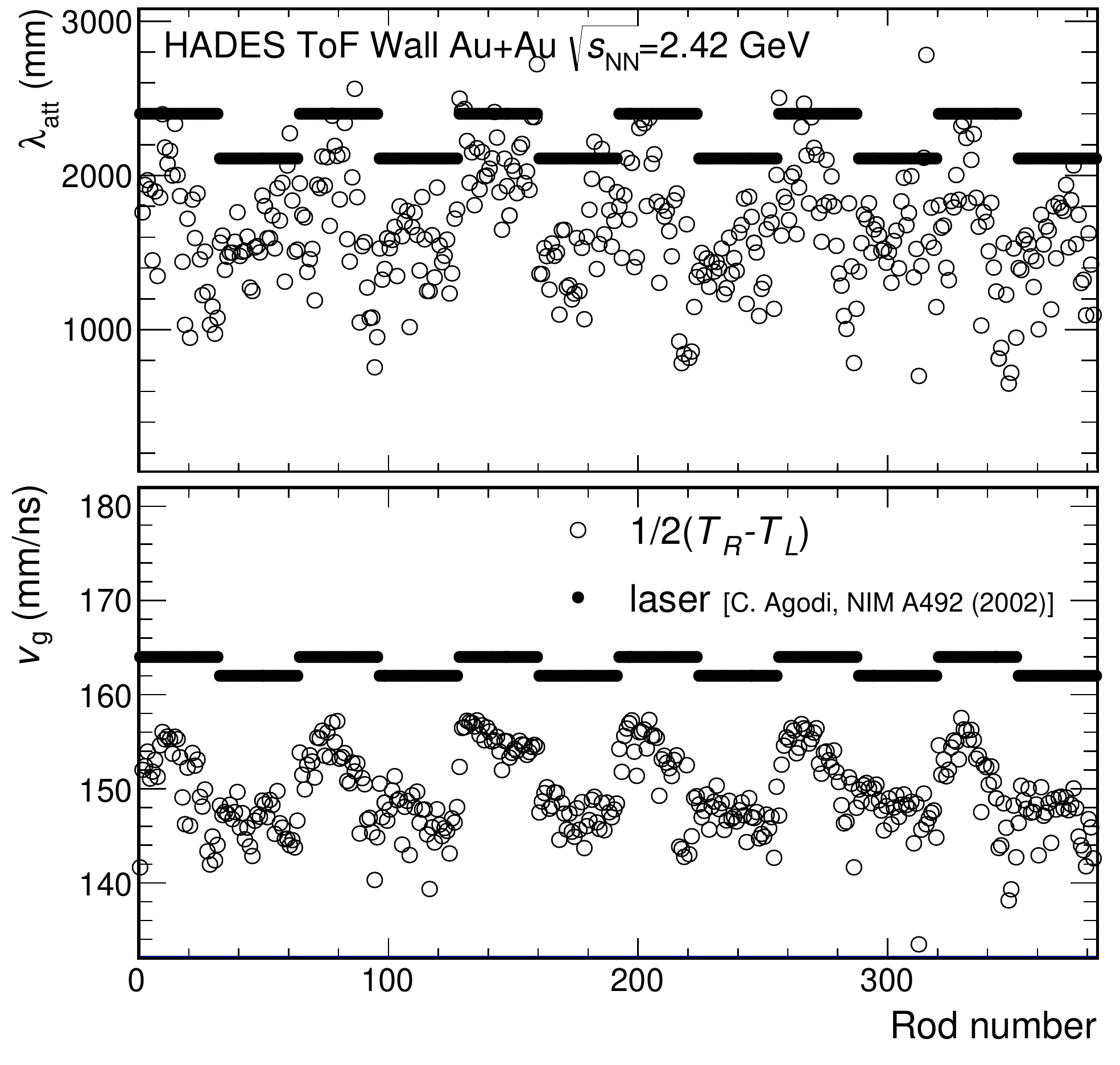}
	\caption{\label{fig_att} Effective attenuation lengths (top panel) and effective group velocity (bottom panel) shown with open circles of all 384 scintillator rods obtained from the data selection. The thick lines are values obtained in a dedicated laser measurement. The rod number $N$ is obtained as $K + 64\times (n_{\rm sector}-1)$, $K$ being the rod number within a sector and $n_{\rm sector}$ the sector number.}
\end{figure}

Clean samples of identified particles for calibration purposes can be prepared considering that in Au + Au collisions at $\sqrt{s_{NN}}=2.42$~GeV protons dominate at all angles and momenta. 
The situation is very different for negatively-charged particles; 
the production of antiprotons is far below the threshold and the majority of negatively-charged particles are pions and electrons. 
Other long-lived particles such as antikaons are strongly suppressed ($\sim10^{-4}~{\rm K}^-$ per event compared to $\sim10~\pi^-$). 
Most of the electrons can be rejected from the sample requiring momenta larger than 200 MeV$/c$, allowing one to attain rather clean data samples of protons and negative pions. 
The purity of the data is additionally enhanced using the energy loss vs momentum correlation measured in the MDC. 
After this last procedure, a purity of about 97~\% of negative pions  and 95~\% of protons is attained, such that the residual contaminations are marginal for the purposes of this study.

The time calibration is a two-step process. 
First, the effective values of $v_{\rm g}$ and $\lambda_{\rm att}$ are obtained from the measured times for each scintillator detector.
In the second step, the correction of the time-walk effect occurs. 
This correction depends on the longitudinal position along the scintillator bar. 
The time walk also compensates for the constant time offset, synchronizing the TOF Wall detector. 
The methods used in both steps are explained in detail below.

\subsection{Determination of the effective attenuation length and effective light group velocity}
\label{subsec:effective_att_vg}

The effective values for the optical attenuation length and light group velocity are obtained from the clean charged particle sample.
The $\lambda_{\rm att}$ of all scintillator rods have been obtained from the slopes of the $\frac{1}{2}\mathtt{ln}\left( \frac{Q_{\rm L}}{Q_{\rm R}}\right)$ distribution as a function of the impact position along each rod. 
Figure~\ref{fig_att} shows a compilation of these slopes for all individual rods compared to the results obtained with laser measurements~\cite{Agodi:Hades_TOF_2002}. 
The light group velocities have been calculated following a similar procedure by fitting the slopes of the $1/2(T_{\rm R}-T_{\rm L})$ vs position distribution. 
The extracted effective values of $v_{\rm g}$ are depicted in Fig.~\ref{fig_att} and clearly show a difference of $\simeq 5$~mm/ns between the thick and thin modules.   
 
These results can be compared to values measured with light pulses from a nitrogen laser injected via a fiber coupler to the rods \cite{Agodi:Hades_TOF_2002} (see Fig.\ref{fig_att}). 
The obtained values were $\lambda_{\rm att} = 2400$~mm for the $30\times30$~mm$^2$ thick rods and $\lambda_{\rm att} = 2100$~mm for the $20\times20$~mm$^2$ thick modules, both with very small dispersion. 
In the same measurement, the light group velocity was found to be 164~mm/ns and 162~mm/ns, respectively, which is larger by 10~mm/ns than the value extracted in this work. 
Hence, the values obtained from the slope analysis should rather be understood as effective values reflecting the response of the whole measurement chain, and not as an intrinsic property of the scintillator itself. 
Additionally, the spread in the values of the effective attenuation length of around 30~\% and of the effective light velocity of 5~\% have been reported in other studies as in ref.~\cite{Carman:2020yma}.

\subsection{Time-walk correction method}

\label{subsec:signal_walk}
\begin{figure}[tb!]
	\centering
	\includegraphics[width=0.47\textwidth]{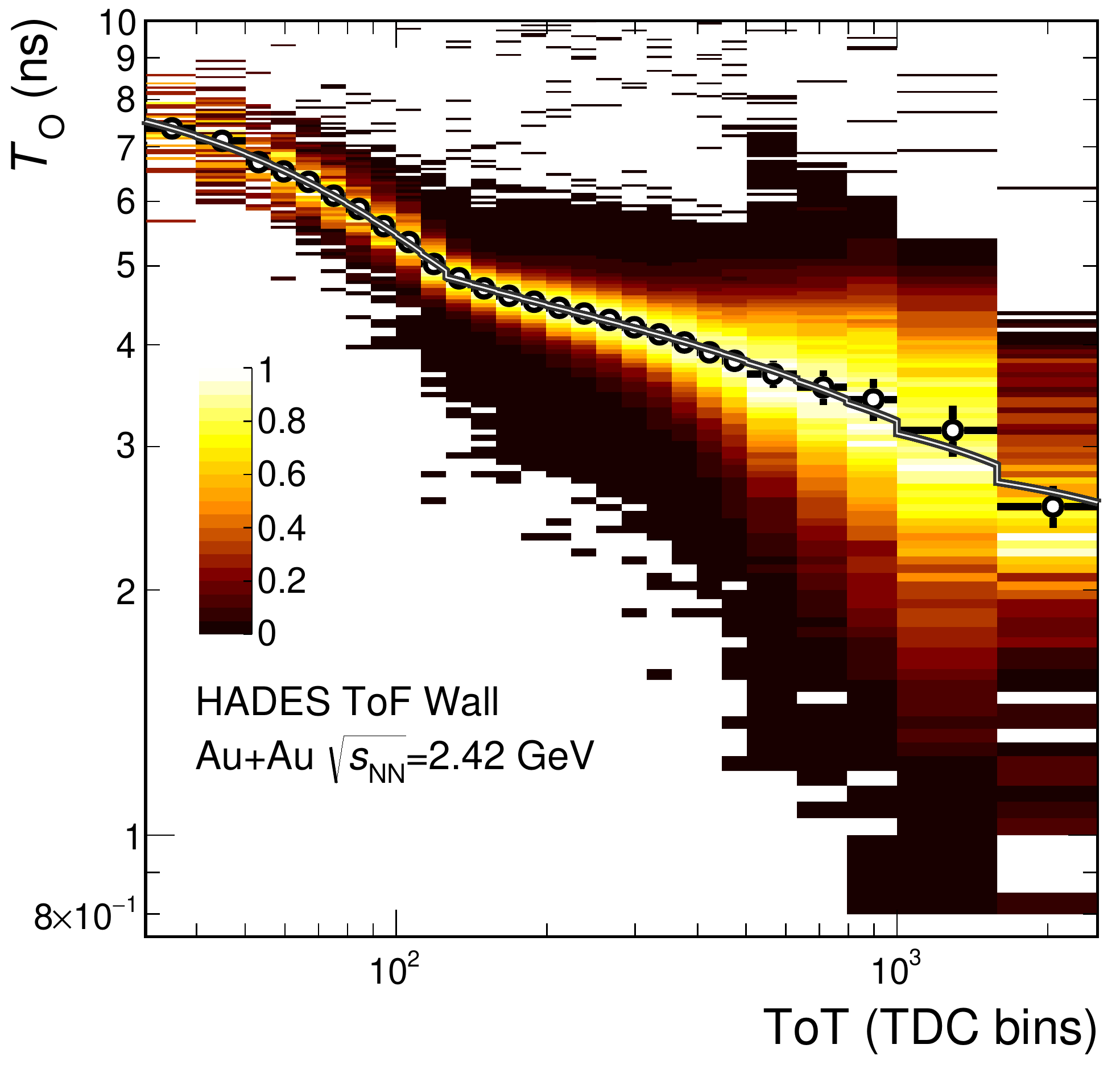}
	\caption{\label{fig_walc_corr_example_same_side_2d} Example of a walk effect correlation. Open symbols depict the median values in each amplitude (ToT) interval. ToT is measured in TDC bin units having a width of 0.098 ns each bin. The black curve is a 9 parameter fit function consisting of an exponential that describes the low ToT region, a linear combination of an exponential and a power-law function describing the high-ToT region and the ToT value which separates into the low and high regions of the spectrum.}
\end{figure}

Signals produced by particles depositing primordial ionization energy in a range from 0.5 to 40 times in MIP equivalent units are efficiently reconstructed. 
Most of the signals having a charge signal below MIP originate from particles crossing the edge of the bar, 
whereas the large $E_{\rm dep}$ corresponds in general to nuclear fragments with Z$>$1, many of which are fully stopped in the scintillator material.  
This spread in $E_{\rm dep}$ leads to ToT values spanning over more than two orders of magnitude. 

The correlation between the registered arrival time of the signal and its ToT can be studied observing the time offset $T_{\rm O}$, which can be obtained as 
\begin{equation}
T_{\rm O} = T_{\rm L/R} - T(p,m) -  \left(\frac{L}{2}\pm x\right)\frac{1}{v_{\rm g}}, 
\end{equation}
where $T(p,m)$, time-of-flight as a function of momentum for the identified particles, was obtained after PID using only the MDC detector as $T(p,m) = D \sqrt{m^2+p^2}/p$. 
Velocity variations along the path due to energy loss and multiple scattering in upstream materials are accounted for by a correction obtained from a MC simulation with realistic underlying events using the UrQMD event generator~\cite{Bass:1998ca}, the full material budget, and the detailed detector response implemented in GEANT3~\cite{Brun:1987ma}. 

An example of such a correlation for a fixed x-position in one rod is shown in  Fig.~\ref{fig_walc_corr_example_same_side_2d}. 
In total, 9 parameters are needed to describe the correction value (time offset). The first 3 parameters describe the low ToT region with an exponential function. The high ToT region is described with a linear combination of an exponential and power-law function with 5 parameters. One additional parameter separates the regions of low and high ToT. 

\begin{figure*}
	\centering
	\includegraphics[width=0.47\textwidth]{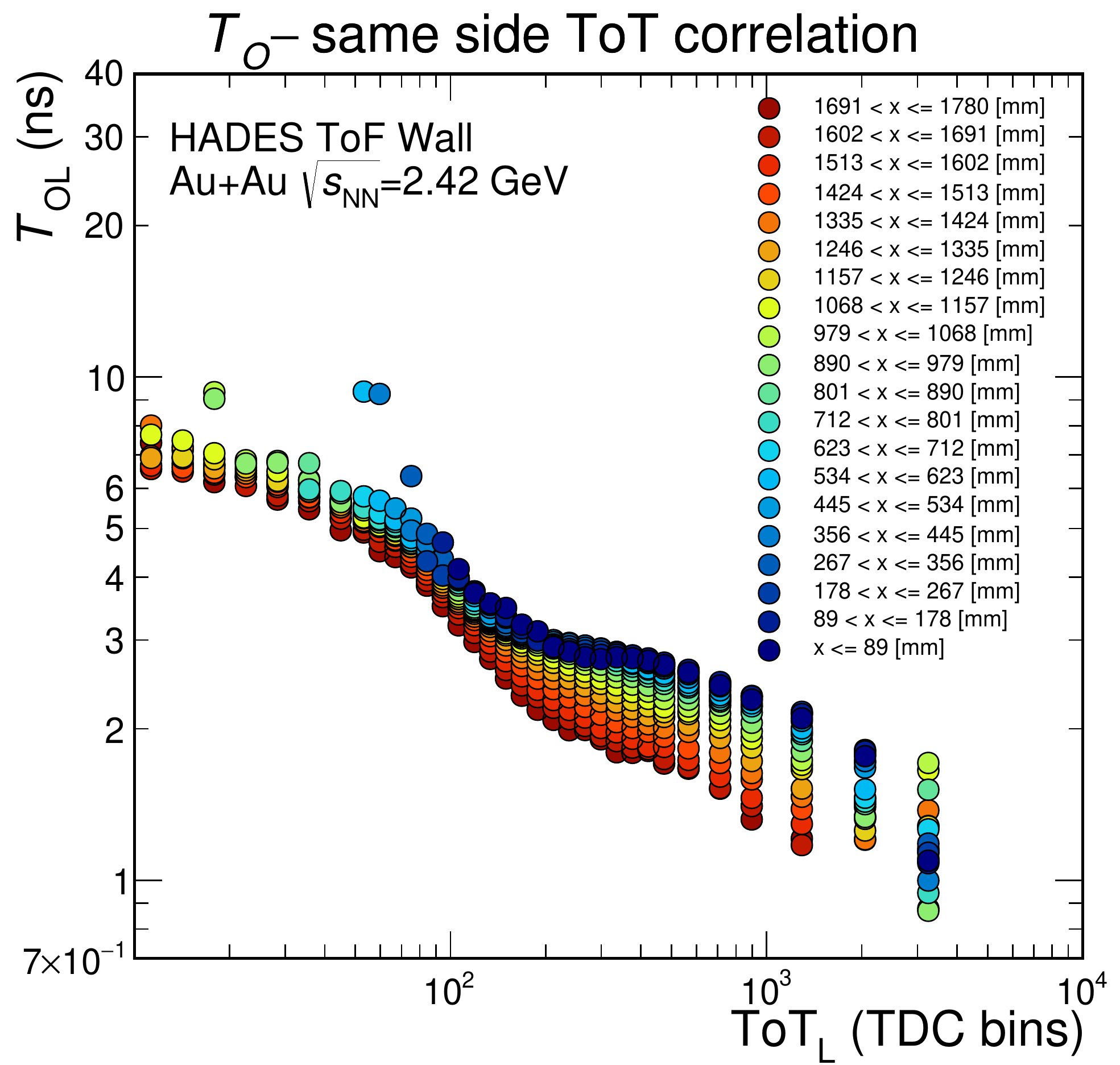}
	\includegraphics[width=0.47\textwidth]{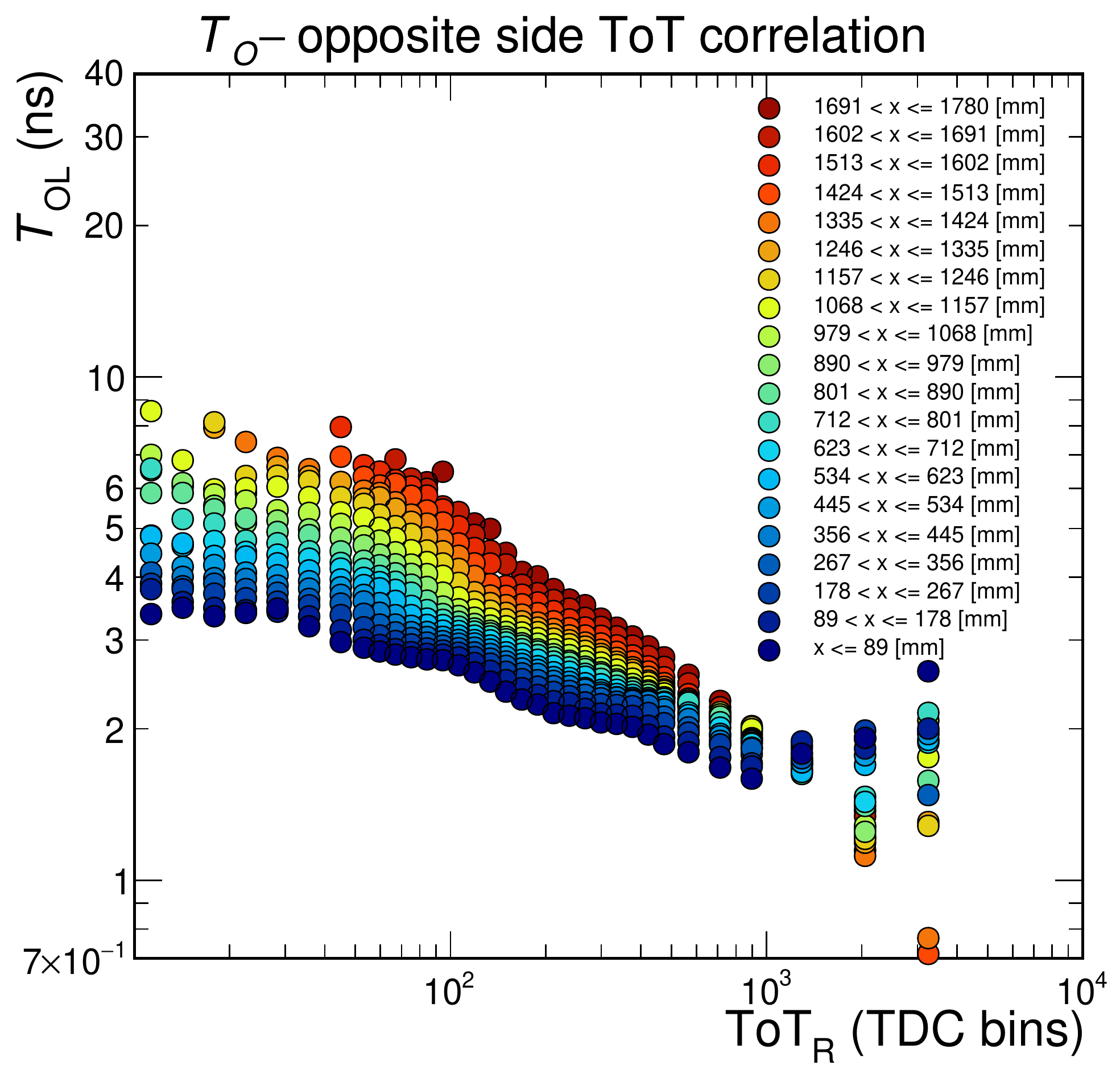}
	\caption{\label{fig_walc_corr_same_side} Walk effect for different longitudinal slices of one scintillator rod. Left: mean values of $T_O$ measured with same side PMT. Right: same measured with opposite side PMT.}
\end{figure*}

To account for the position dependence of the time walk, the time offset--ToT dependence is obtained for 20 equal segments along each scintillator. 
Within a segment, the signal attenuation can be assumed to be constant. 
Hence, the correlation between the time offset and the amplitude can also be obtained with respect to the amplitude measured at the opposite side of the scintillator bar. 
The corresponding correlations for the same and opposite side readouts of all 20 longitudinal bins are shown in Figure~\ref{fig_walc_corr_same_side}. 
The time offsets range from 1 up to 10 ns and constitute a significant fraction of the total time, comparable in magnitude to the flight time. 
The walk time value is obtained using a weighted average of the same and opposite side offsets. 
The real values $T_{\rm R/L}$ are then obtained by subtracting the walk time from the raw times $T^{\rm raw}_{\rm R/L}$ as 
\begin{equation}
T_{\rm R/L} = T^{\rm raw}_{\rm R/L} - \frac{f(ToT_{\rm R/L})^{\rm s}\omega_{\rm s} + f(ToT_{\rm L/R})^{\rm o}\omega_{\rm o} }{\omega_{\rm s}\omega_{\rm o}}, 
\end{equation}  
where the indices s and o denote the same and opposite side time-ToT correlations and $\omega_{\rm s/o}$ the respective weights. 
The weights have been obtained from the spread of the distributions observed in the real data. 

%%%%%%%%%%%%%%%%%%%%%%%%%%%%%%%%
%%%%%%%%%%%%%%%%%%%%%%%%%%%%%%%%
%%%%%%%% PERFORMANCE %%%%%%%%%%%
%%%%%%%%%%%%%%%%%%%%%%%%%%%%%%%%
%%%%%%%%%%%%%%%%%%%%%%%%%%%%%%%%
\section{Detector performance}
\label{sec:performance}

\begin{figure}[tb!]
	\centering
	\includegraphics[width=0.47\textwidth]{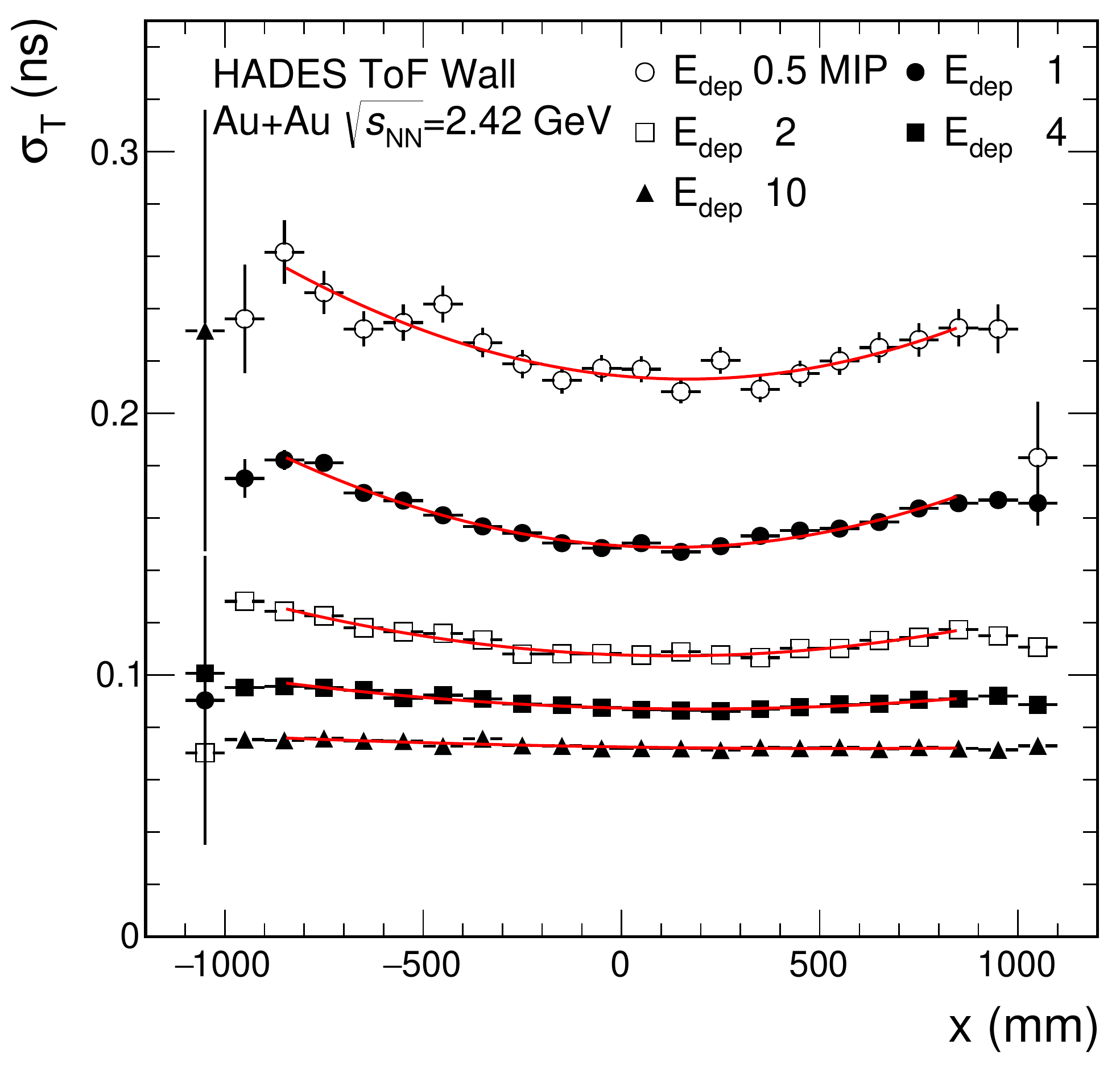}
	\caption{\label{fig_fit_cosh_two_sides}Time of flight uncertainty of a typical rod calculated from position residuals as a function of longitudinal position along the rod for different values of deposited energy in the scintillator $E_{dep}$. The 0.5 MIP equivalent corresponds to edge hits.}
\end{figure}

\begin{figure}[tb!]
	\includegraphics[width=0.47\textwidth]{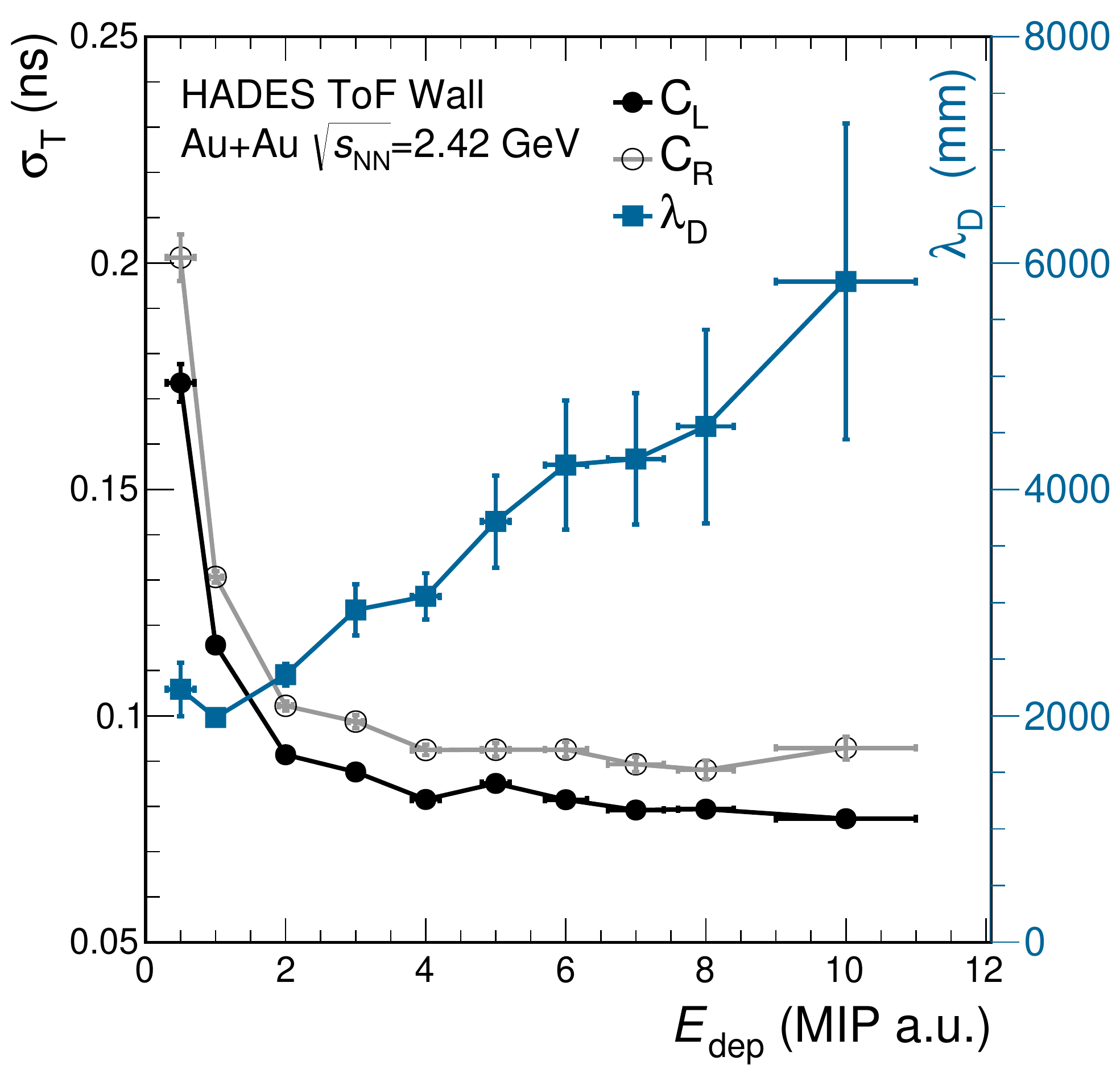}
	\caption{\label{fig_effective_attenuations_from_cosh_fits} Individual time accuracy constants for left and right PMT readout (left axis) and
 the timing degradation length (right axis) obtained from the position measurement as a function of the deposited energy $E_{\rm dep}$.}
\end{figure}

For the analysis of the timing performance as a function of the primordial deposited energy in the scintillator, we have selected pion tracks with a range of deposited energy in the scintillator starting from half the value of a MIP up to its 10 fold value. 
The uncertainty of the measured TOF when using eq.~\ref{eq:resolution_left_right_time}, is
\begin{equation}
\sigma_{\rm T} = \left( \sigma_{T_{\rm L}}^2 + \sigma_{T_{\rm R}}^2\right)^{1/2}.\label{eq:total_sigma_tof}
\end{equation}
The best timing performance is found at the center of the scintillator rod and the worst happens at the edges. 

The position and time uncertainties have both identical distributions, but the first one is scaled by $v_{\rm g}$. 
However, the position is better determined by MDC than by TOF. 
In the latter case, when characterizing the variance of the $T-T(p,m)$ distribution, the contributions from the START, multiple scattering and flight path uncertainties have to be accounted for and properly subtracted. 
The position uncertainty is determined by the variance of the distribution $x-x_{\rm MDC}$, where $x_{\rm MDC}$ is pointing at the scintillator bar from the MDC tracking stations. 
Since this pointing accuracy is at least one order of magnitude better than the resulting variance of $x$, its contribution to the final uncertainty is negligible. 

The $\sigma_T$ values for one typical scintillator bar are depicted in Fig.~\ref{fig_fit_cosh_two_sides} along x for 5 different ranges in $E_{\rm dep}$. 
The measured timing performance in each $E_{\rm dep}$ range is fitted with Eq.~\ref{eq:total_sigma_tof} to obtain the effective degradation length $\lambda_{\rm D}$ as well as the left and right constants, $C_{\rm L}$ and $C_{\rm R}$, respectively. 
The trends of these three values are shown in Fig.~\ref{fig_effective_attenuations_from_cosh_fits}. 
The $\lambda_{\rm D}$ increases from 2000~mm up to 6000~mm almost linearly with increasing signal amplitude. 
The left and right side constants decrease from approximately 0.2 ns for pions depositing 0.5 MIP equivalent energy to 0.08-0.09 ns when reaching 4 times MIP. 
Above that value it stays constant. 
This plateau could be interpreted as reaching the limits of the channel TDC chain performance.

\begin{figure}[tb!]
	\centering
	\includegraphics[width=0.47\textwidth]{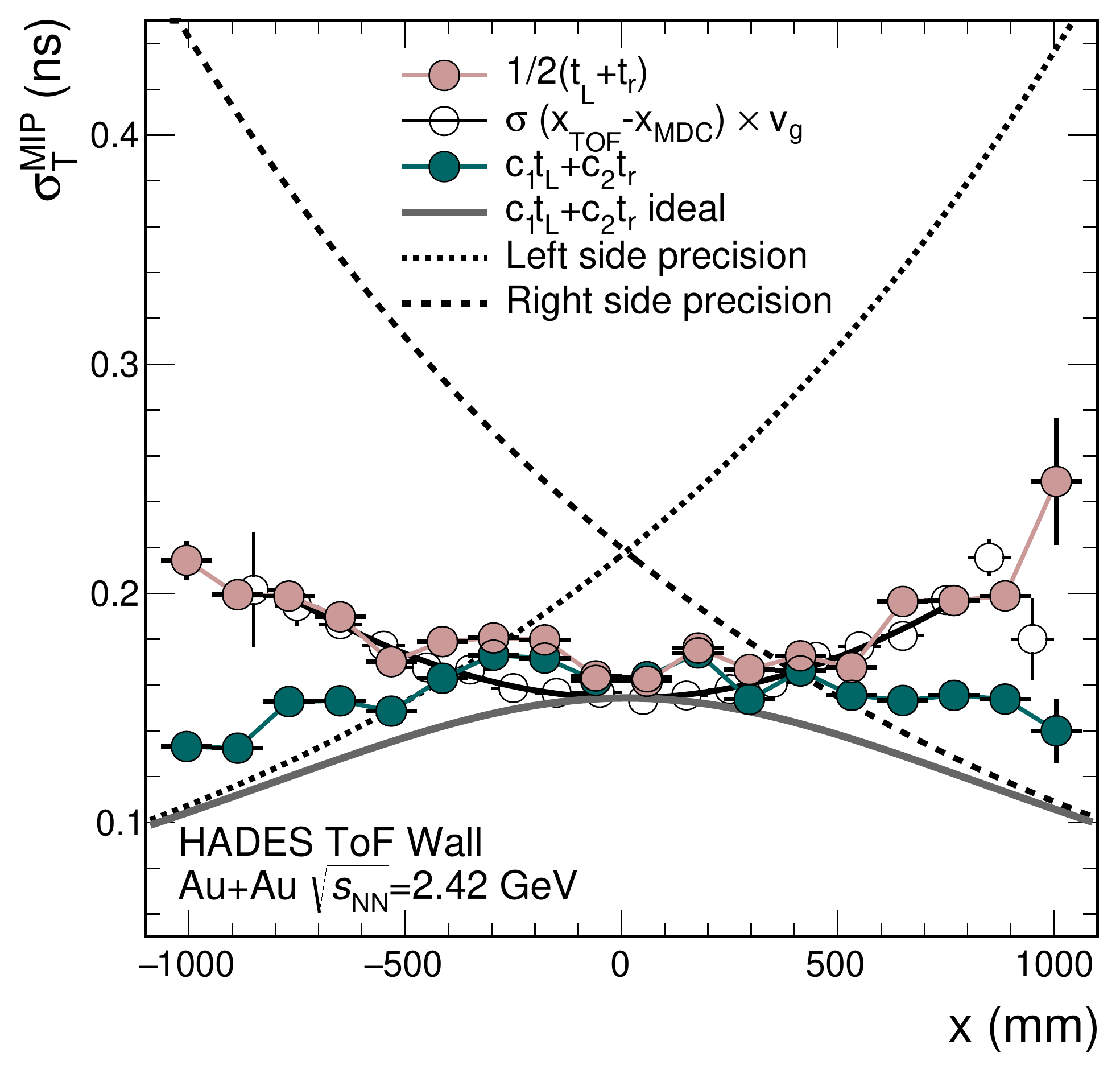}
	\caption{\label{fig_fit_resolution_from_tof_mips} Time of flight uncertainties calculated for negative MIP pions (momentum between 400 and 450 MeV/$c$) using the unweighted average of left and right times (red circles) compared to the values for the weighted averages (green circles). The open symbols denote uncertainties calculated from position residuals. The dotted and dashed lines show the contributions of each individual side and the thick grey line depicts the expectation from position measurement with the weighted method and without additional uncertainties.}
\end{figure}

The comparison of the timing performance obtained with the weighted average using Eq.~\ref{eq:tof_weighted} and the unweighted average using Eq.~\ref{eq:tof} for MIPs is shown in Figure \ref{fig_fit_resolution_from_tof_mips}.  
The additional contribution originating from the START time was evaluated and is 54~ps \cite{Kornakov:GSI16Report}. 
The values obtained for the unweighted case coincide with the expectation from the position analysis, shown with open circles. 
The weighted average results (green circles) coincide at the rod center but improve significantly towards the edges and are 10-15 ps higher than the values expected from the position analysis, shown with a grey thick line. 
The curve was obtained considering the case when no additional sources contribute to the smearing of the time signal.
However, if adding quadratically the START time uncertainty, the expected value is increased by 15~ps at the edges of the scintillator and by 10~ps in the middle, improving the quantitative agreement.  

\begin{figure}[tb!]
	\centering
	\includegraphics[width=0.47\textwidth]{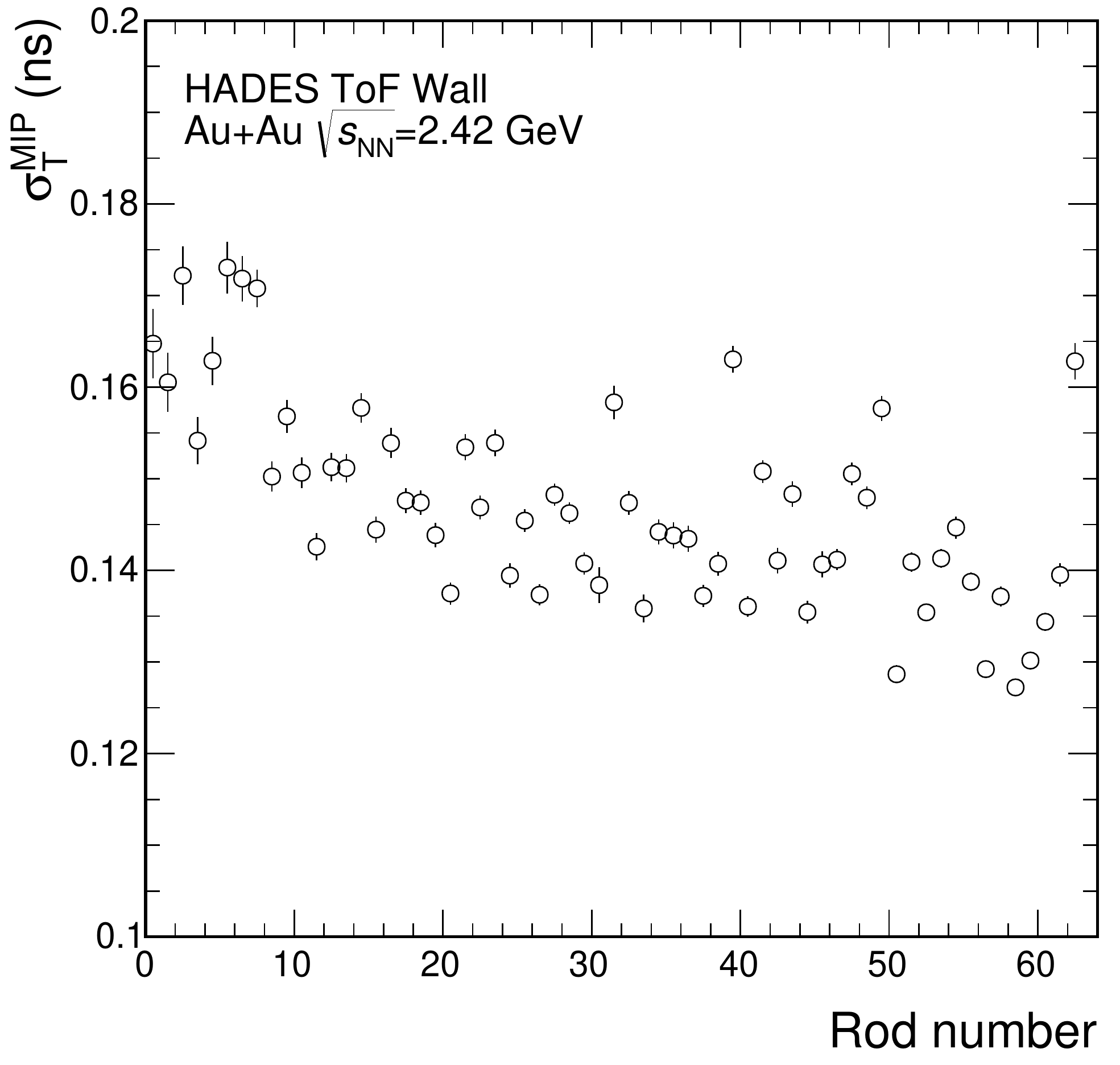}
	\caption{\label{fig_resolution_pion_mips_rod_number} Mean time of flight uncertainty for MIPs in one sector using the weighted average method. The most significant contribution to the falling trend is due to the rod lengths.}
\end{figure}

The average time precision for MIPs is shown for all rods in Fig.~\ref{fig_resolution_pion_mips_rod_number}. 
The values show a decreasing trend with rod number, i.e with decreasing scintillator length. The values of 160~ps and of 135~ps for the longest and shortest rods, respectively, are are in accordance with previous laser measurements \cite{Agodi:Hades_TOF_2002}. 
This difference is mainly due to the higher average degradation of the time signal in the longest rods. 
The spread of values for rods of equal length is of the order of 15~ps. 
The averaged timing performance of the TOF Wall detector is about 150~ps. This number has to be compared to 190~ps which is attained when the position dependence of the time walk is not considered. 

\begin{figure}[tb!]
	\centering
	\includegraphics[width=0.47\textwidth]{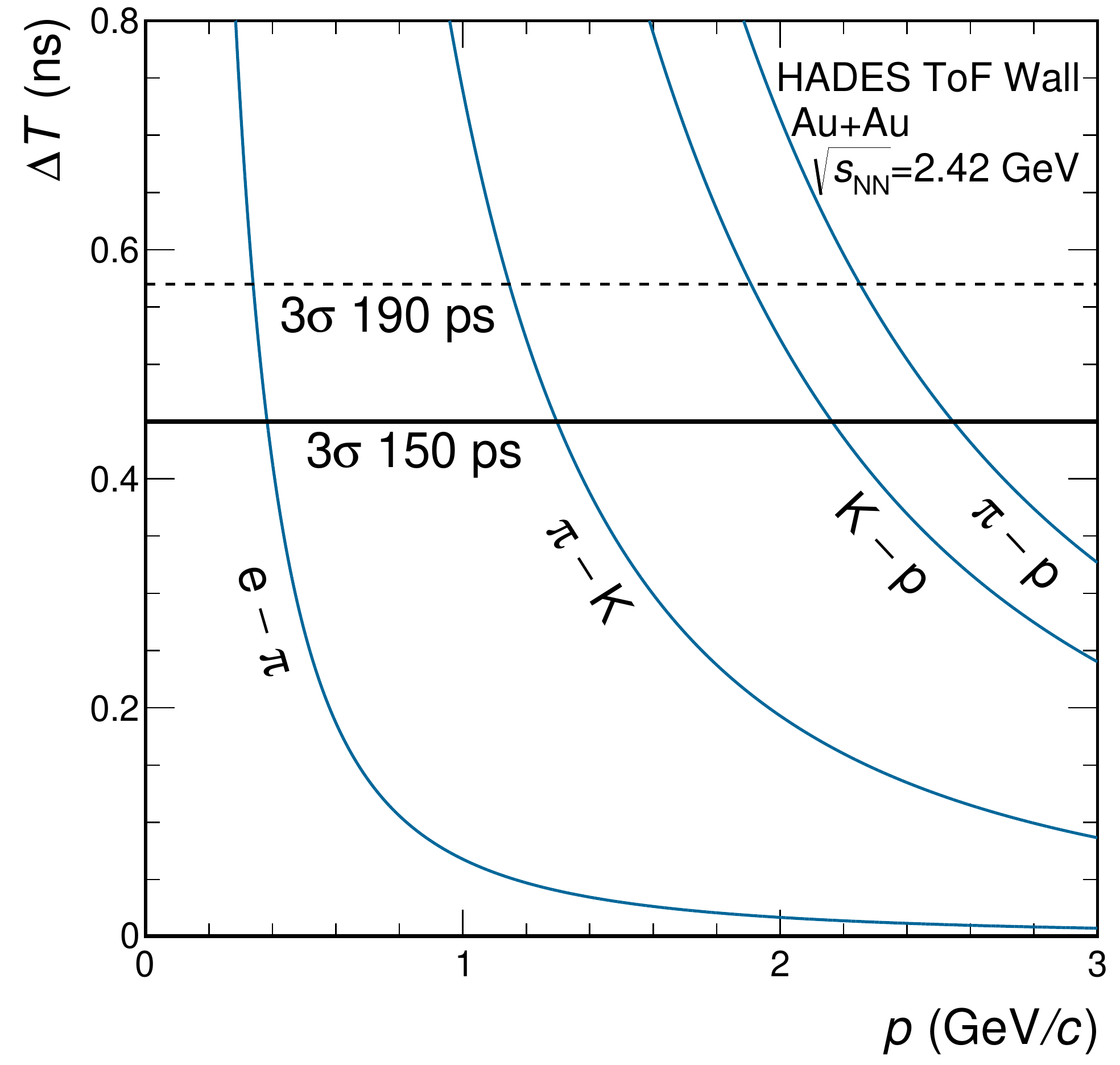}
	\caption{\label{fig_separation_power_sigmas} The time of flight differences for light charged particles as a function of momentum and 2100 mm of the flight path in the experiment. The solid line indicates the 3-$\sigma$ region when the position dependence of the time-walk is considered and the long dashed line corresponds to the case when this effect is neglected and the time-walk correction consist in one single function for each scintillator. }
\end{figure}

The timing performance can be interpreted in terms of particle separation power, as shown in Fig.~\ref{fig_separation_power_sigmas}. 
The achieved average timing performance of the TOF Wall improves the separation of electrons from pions at the 3-$\sigma$ level up to almost 400~MeV$/c$, pions and kaons up to 1400~MeV$/c$ and kaons from protons up to 2200~MeV$/c$ considering an average flight distance of 2100~mm. 

\section{Summary and conclusions}
\label{sec_discussion}

In this work, we describe a calibration and time-walk correction  method utilizing the  position-dependent time-ToT correlation. 
The standard correction in the form of an additional polynomial, as implemented in~\cite{Kichimi:2000uu,Wu:2005xk,Denisov:GlueX_position_dependent_walk_2002,Carman:2020yma},  was not sufficient and the signal template method~\cite{Moskal:2014rja} was not applicable since the full pulse trace is not measured. 
The main idea followed here is to split the long scintillator rod into regions with approximately constant signal attenuation and determine the time offset--ToT dependence. We find that 20 longitudinal segments of approximately 7-12 cm length resolve the position dependence problem. Larger segmentation does not result in increased performance or improved timing accuracy.

The timing response is evaluated with two different methods. 
The first one uses the precise pointing of the tracking system to measure the position uncertainty and evaluate from it the timing performance. 
In this approach, contributions from the START time, multiple scattering effects, or energy loss are not included. 
The second method evaluates directly the time uncertainty from a clean sample of identified negative pion tracks. 
The second method is compatible with expectations if the START detector precision is included.

The average time uncertainty for MIP pions (momentum between 400 and 450~MeV/$c$) is found to be 130~ps in the shortest rods of 1475~mm and 165~ps in the longest of 2365~mm. 
The overall average performance improves from 190~ps to 150~ps when applying position-dependent time-walk correction. This improvement allows in the HADES energy regime to better discriminate between different particle species at the 3-$\sigma$ level. In the case of electrons and pions, it increases the range by 40~MeV$/c$, for pions and kaons it is improved by 150~MeV$/c$, and between kaons and protons by 250~MeV$/c$.

This work has been supported by Warsaw University of Technology, Warsaw (Poland): OPUS grant from National Science Center of Poland 2017/27/B/ST2/01947;
TU Darmstadt, Darmstadt (Germany): VH-NG-823, DFG GRK 2128, DFG CRC-TR 211, BMBF:05P18RDFC1; NPI CAS, Rez, Rez (Czech Republic): MSMT LM2018112, OP VVV CZ.02.1.01/0.0/0.0/18\_046/0016066, LTT17003.

\section*{References}

\bibliography{mybibfile}

\begin{thebibliography}{21}
\expandafter\ifx\csname natexlab\endcsname\relax\def\natexlab#1{#1}\fi
\providecommand{\url}[1]{\texttt{#1}}
\providecommand{\href}[2]{#2}
\providecommand{\path}[1]{#1}
\providecommand{\DOIprefix}{doi:}
\providecommand{\ArXivprefix}{arXiv:}
\providecommand{\URLprefix}{URL: }
\providecommand{\Pubmedprefix}{pmid:}
\providecommand{\doi}[1]{\href{http://dx.doi.org/#1}{\path{#1}}}
\providecommand{\Pubmed}[1]{\href{pmid:#1}{\path{#1}}}
\providecommand{\bibinfo}[2]{#2}
\ifx\xfnm\relax \def\xfnm[#1]{\unskip,\space#1}\fi
%Type = Article
\bibitem[{Tsujita et~al.(1996)Tsujita, Asano, Hamasaki, Mori, Yusa, and
  Kephart}]{Tsujita:1996tk}
\bibinfo{author}{T.~Tsujita}, \bibinfo{author}{Y.~Asano},
  \bibinfo{author}{H.~Hamasaki}, \bibinfo{author}{S.~Mori},
  \bibinfo{author}{K.~Yusa}, \bibinfo{author}{R.~D. Kephart},
  \bibinfo{journal}{Nucl. Instrum. Meth. A} \bibinfo{volume}{383}
  (\bibinfo{year}{1996}) \bibinfo{pages}{413--423}.
  \DOIprefix\doi{10.1016/S0168-9002(96)00871-6}.
%Type = Article
\bibitem[{Kichimi et~al.(2000)}]{Kichimi:2000uu}
\bibinfo{author}{H.~Kichimi}, et~al., \bibinfo{journal}{Nucl. Instrum. Meth. A}
  \bibinfo{volume}{453} (\bibinfo{year}{2000}) \bibinfo{pages}{315--320}.
  \DOIprefix\doi{10.1016/S0168-9002(00)00651-3}.
%Type = Article
\bibitem[{Denisov et~al.(2002)}]{Denisov:GlueX_position_dependent_walk_2002}
\bibinfo{author}{S.~Denisov}, et~al., \bibinfo{journal}{Nucl. Instrum. Meth.}
  \bibinfo{volume}{A494} (\bibinfo{year}{2002}) \bibinfo{pages}{495--499}.
  \DOIprefix\doi{10.1016/S0168-9002(02)01538-3}.
%Type = Article
\bibitem[{Agodi et~al.(2002)}]{Agodi:Hades_TOF_2002}
\bibinfo{author}{C.~Agodi}, et~al., \bibinfo{journal}{Nucl. Instrum. Meth.}
  \bibinfo{volume}{A492} (\bibinfo{year}{2002}) \bibinfo{pages}{14--25}.
  \DOIprefix\doi{10.1016/S0168-9002(02)01004-5}.
%Type = Article
\bibitem[{Wu et~al.(2005)}]{Wu:2005xk}
\bibinfo{author}{C.~Wu}, et~al., \bibinfo{journal}{Nucl. Instrum. Meth. A}
  \bibinfo{volume}{555} (\bibinfo{year}{2005}) \bibinfo{pages}{142--147}.
  \DOIprefix\doi{10.1016/j.nima.2005.09.029}.
%Type = Article
\bibitem[{Moskal et~al.(2015)}]{Moskal:2014rja}
\bibinfo{author}{P.~Moskal}, et~al., \bibinfo{journal}{Nucl. Instrum. Meth. A}
  \bibinfo{volume}{775} (\bibinfo{year}{2015}) \bibinfo{pages}{54--62}.
  \DOIprefix\doi{10.1016/j.nima.2014.12.005}.
  \href{http://arxiv.org/abs/1412.6963}{\tt arXiv:1412.6963}.
%Type = Article
\bibitem[{Carman et~al.(2020)}]{Carman:2020fsv}
\bibinfo{author}{D.~Carman}, et~al., \bibinfo{journal}{Nucl. Instrum. Meth. A}
  \bibinfo{volume}{960} (\bibinfo{year}{2020}) \bibinfo{pages}{163629}.
  \DOIprefix\doi{10.1016/j.nima.2020.163629}.
%Type = Article
\bibitem[{Agakishiev et~al.(2009)}]{Agakishiev:Hades_Overwiew_2009}
\bibinfo{author}{G.~Agakishiev}, et~al. (\bibinfo{collaboration}{HADES}),
  \bibinfo{journal}{Eur. Phys. J.} \bibinfo{volume}{A41} (\bibinfo{year}{2009})
  \bibinfo{pages}{243--277}. \DOIprefix\doi{10.1140/epja/i2009-10807-5}.
  \href{http://arxiv.org/abs/0902.3478}{\tt arXiv:0902.3478}.
%Type = Article
\bibitem[{Michel et~al.(2011)Michel, {B\"ohmer}, Kajetanowicz, Korcyl, Maier,
  Palka, Stroth, Tarantola, Traxler, Ugur, and Yurevich}]{TRBCollaboration}
\bibinfo{author}{J.~Michel}, \bibinfo{author}{M.~{B\"ohmer}},
  \bibinfo{author}{M.~Kajetanowicz}, \bibinfo{author}{G.~Korcyl},
  \bibinfo{author}{L.~Maier}, \bibinfo{author}{M.~Palka},
  \bibinfo{author}{J.~Stroth}, \bibinfo{author}{A.~Tarantola},
  \bibinfo{author}{M.~Traxler}, \bibinfo{author}{C.~Ugur},
  \bibinfo{author}{S.~Yurevich}, \bibinfo{journal}{Journal of Instrumentation}
  \bibinfo{volume}{6} (\bibinfo{year}{2011}) \bibinfo{pages}{C12056}.
  \URLprefix \url{http://stacks.iop.org/1748-0221/6/i=12/a=C12056}.
%Type = Article
\bibitem[{Kim et~al.(2015)}]{Kim:2015tva}
\bibinfo{author}{S.~Kim}, et~al., \bibinfo{journal}{Nucl. Instrum. Meth. A}
  \bibinfo{volume}{795} (\bibinfo{year}{2015}) \bibinfo{pages}{39--44}.
  \DOIprefix\doi{10.1016/j.nima.2015.05.046}.
%Type = Article
\bibitem[{Denniston et~al.(2020)}]{Denniston:2020gmc}
\bibinfo{author}{A.~Denniston}, et~al., \bibinfo{journal}{Nucl. Instrum. Meth.
  A} \bibinfo{volume}{973} (\bibinfo{year}{2020}) \bibinfo{pages}{164177}.
  \DOIprefix\doi{10.1016/j.nima.2020.164177}.
  \href{http://arxiv.org/abs/2004.10268}{\tt arXiv:2004.10268}.
%Type = Article
\bibitem[{Carman et~al.(2020)Carman, Asryan, Baturin, Clark, De~Vita, Kim,
  Miller, and Wiggins}]{Carman:2020yma}
\bibinfo{author}{D.~Carman}, \bibinfo{author}{G.~Asryan},
  \bibinfo{author}{V.~Baturin}, \bibinfo{author}{L.~Clark},
  \bibinfo{author}{R.~De~Vita}, \bibinfo{author}{W.~Kim},
  \bibinfo{author}{B.~Miller}, \bibinfo{author}{C.~Wiggins},
  \bibinfo{journal}{Nucl. Instrum. Meth. A} \bibinfo{volume}{960}
  (\bibinfo{year}{2020}) \bibinfo{pages}{163626}.
  \DOIprefix\doi{10.1016/j.nima.2020.163626}.
%Type = Article
\bibitem[{Moskal et~al.(2020)}]{Moskal:2020jrl}
\bibinfo{author}{P.~Moskal}, et~al., \bibinfo{journal}{IEEE Trans. Instrum.
  Measur.} \bibinfo{volume}{70} (\bibinfo{year}{2020}) \bibinfo{pages}{1--10}.
  \DOIprefix\doi{10.1109/TIM.2020.3018515}.
  \href{http://arxiv.org/abs/2008.10868}{\tt arXiv:2008.10868}.
%Type = Article
\bibitem[{Adamczewski-Musch et~al.(2018)}]{Adamczewski-Musch:2017sdk}
\bibinfo{author}{J.~Adamczewski-Musch}, et~al.
  (\bibinfo{collaboration}{HADES}), \bibinfo{journal}{Eur. Phys. J. A}
  \bibinfo{volume}{54} (\bibinfo{year}{2018}) \bibinfo{pages}{85}.
  \DOIprefix\doi{10.1140/epja/i2018-12513-7}.
  \href{http://arxiv.org/abs/1712.07993}{\tt arXiv:1712.07993}.
%Type = Article
\bibitem[{Kornakov et~al.(2014)}]{Kornakov:2014cua}
\bibinfo{author}{G.~Kornakov}, et~al. (\bibinfo{collaboration}{HADES}),
  \bibinfo{journal}{JINST} \bibinfo{volume}{9} (\bibinfo{year}{2014})
  \bibinfo{pages}{C11015}. \DOIprefix\doi{10.1088/1748-0221/9/11/C11015}.
%Type = Article
\bibitem[{Anghinolfi et~al.(2004)Anghinolfi, Jarron, Martemyanov, Usenko,
  Wenninger, Williams, and Zichichi}]{Anghinolfia:NINO2004}
\bibinfo{author}{F.~Anghinolfi}, \bibinfo{author}{P.~Jarron},
  \bibinfo{author}{A.~N. Martemyanov}, \bibinfo{author}{E.~Usenko},
  \bibinfo{author}{H.~Wenninger}, \bibinfo{author}{M.~C.~S. Williams},
  \bibinfo{author}{A.~Zichichi}, \bibinfo{journal}{Nucl. Instrum. Meth.}
  \bibinfo{volume}{A533} (\bibinfo{year}{2004}) \bibinfo{pages}{183--187}.
  \DOIprefix\doi{10.1016/j.nima.2004.07.024}.
%Type = Article
\bibitem[{Pietraszko et~al.(2014)Pietraszko, Galatyuk, Grilj, Koenig, Spataro,
  and {Tr\"ager}}]{START}
\bibinfo{author}{J.~Pietraszko}, \bibinfo{author}{T.~Galatyuk},
  \bibinfo{author}{V.~Grilj}, \bibinfo{author}{W.~Koenig},
  \bibinfo{author}{S.~Spataro}, \bibinfo{author}{M.~{Tr\"ager}}
  (\bibinfo{collaboration}{HADES}), \bibinfo{journal}{Nucl. Instrum. Meth.}
  \bibinfo{volume}{A763} (\bibinfo{year}{2014}) \bibinfo{pages}{1--5}.
  \DOIprefix\doi{10.1016/j.nima.2014.06.006}.
%Type = Article
\bibitem[{Kurata et~al.(1994)}]{Kurata:time_degradation_1994}
\bibinfo{author}{M.~Kurata}, et~al., \bibinfo{journal}{Nucl. Instrum. Meth.}
  \bibinfo{volume}{A349} (\bibinfo{year}{1994}) \bibinfo{pages}{447--453}.
  \DOIprefix\doi{10.1016/0168-9002(94)91209-2}.
%Type = Article
\bibitem[{Bass et~al.(1998)}]{Bass:1998ca}
\bibinfo{author}{S.~Bass}, et~al., \bibinfo{journal}{Prog. Part. Nucl. Phys.}
  \bibinfo{volume}{41} (\bibinfo{year}{1998}) \bibinfo{pages}{255--369}.
  \DOIprefix\doi{10.1016/S0146-6410(98)00058-1}.
  \href{http://arxiv.org/abs/nucl-th/9803035}{\tt arXiv:nucl-th/9803035}.
%Type = Article
\bibitem[{Brun et~al.(1987)Brun, Bruyant, Maire, McPherson, and
  Zanarini}]{Brun:1987ma}
\bibinfo{author}{R.~Brun}, \bibinfo{author}{F.~Bruyant},
  \bibinfo{author}{M.~Maire}, \bibinfo{author}{A.~McPherson},
  \bibinfo{author}{P.~Zanarini}  (\bibinfo{year}{1987}).
%Type = Article
\bibitem[{Kornakov(2016)}]{Kornakov:GSI16Report}
\bibinfo{author}{G.~Kornakov} (\bibinfo{collaboration}{HADES}),
  \bibinfo{journal}{GSI Report 2017-1}  (\bibinfo{year}{2016})
  \bibinfo{pages}{109}. \DOIprefix\doi{10.15120/GSI-2017-00527}.

\end{thebibliography}

\end{document}